# Design and implementation of the Customer Experience Data Mart in the Telecommunication Industry: Application Order-To-Payment end to end process

Mounire Benhima[1], John P. Reilly[2], Zaineb Naamane[3], Meriam Kharbat[4], Mohammed Issam Kabbaj[5] and Oussama Esqalli[6]

[1] Eng., M.Eng, Member of TM Forum Trainers Panel / Business Transformation Subject Matter Expert, POWERACT Consulting
Casablanca, Morocco

[2] Subject Matter Expert and Senior Program Manager, Member of TM Forum Trainers Panel, TM Forum
Morristown, NJ, USA

[3] Computer Science Department, Ecole Mohammadia d'ingénieurs, Mohammed V University,
Rabat, Morocco

[4] Computer Science Department, Ecole Mohammadia d'ingénieurs, Mohammed V University,
Rabat, Morocco

[5] Computer Science Department, Ecole Mohammadia d'ingénieurs, Mohammed V University,
Rabat, Morocco

[6] IT / Business Process Management Consultant, MegaTIC Company
Casablanca, Morocco

**Abstract**

Facing the new market challenges, service providers are looking for solutions to improve three major business areas namely the Customer Experience, the Operational Efficiency and Revenue and Margin. To meet the business requirements related to these areas, service providers are going through three major transformation programs namely the Business Support Systems (BSS) transformation program for Customer related aspects, the Operations Support System (OSS) transformation program for mainly service Fulfillment and Assurance and Resource (Application, Computing and Network) Fulfillment and Assurance, and Time To Market (TTM) Transformation program for Products and Services development and management. These transformations are about making a transition from a current situation with all its views (Ex.: Process, Information, Application, Technology) to a desired one. The information view transformation is about reorganizing and reengineering the existing information to be used for the day to day activities and reporting to support decision making. For reporting purpose, service providers have to invest in Business Intelligence (BI) solutions.  for which the main purpose is to provide the right information in a timely manner to efficiently support the decision making. One of the key BI challenges is to model an information structure where to host all the information coming from multiple sources.

The purpose of this paper is to suggest a step by step methodology to design a Telco Data Mart, one of the fundamental BI components. Order-To-Payment, an end to end customer process, will be used as an application for this methodology. Our methodology consists on bringing together the concepts of business intelligence and the telecom business frameworks developed by the TM Forum: the Business Process Framework (eTOM), the Information Framework (SID), and the Business Metrics. Throughout this paper we will highlight the different tasks leading to the implementation of the Telco BI solution, including the mapping between the different telecom frameworks which aim was to refine the work already done by the TM Forum members and to create the missing links between the three frameworks, selecting the convenient performance and quality indicators (KPI), modeling the Customer Experience Data Mart and elaborating the dashboard system. The advantage of this solution is its ability to adapt to any telecom enterprise architecture since it's built around the business standards.

# 1. BI fundamentals

Business Intelligence is a set of concepts, methods, and technologies designed to pursue the elusive goal of turning all the widely separated data in an organization into useful information and eventually into knowledge. (John C. Hancock, 2006)

Business Intelligence revolves around providing easy access to the Information in order to enable the best decision-making in the right moment. This is why, it is necessary to create a separate database with a design and an operational approach that is optimized for queries rather than atomic transactions, this is the data warehouse.

Over time, the various approaches to designing a database schema that is optimized for understanding and querying information have been consolidated into an approach called a dimensional model. At the center of the dimensional model are the numeric measures that we are interested in understanding, such as Mean duration to fulfill service order and % orders delivered by committed date. Related measures are collected into fact tables that contain columns for each of the numeric measures. The most used dimensional model organization is the star schema where we find one or many fact tables in the center, joined to dimension tables. (John C. Hancock, 2006)

## 1.1 BI Architecture

In general, a BI project involves the following steps, as shown in Figure 1:

- **Data Transformation (ETL):** this step consists of extracting, transforming and loading the data from different data sources to the data warehouse.
- **Data analysis (OLAP):** this step provides all the necessary mechanisms to analyze data through different analysis axes.
- **Reporting:** this is the final step of a BI project. It consists on representing a global view of the business activity to the end users using analysis graphs, reports, dashboard, etc.

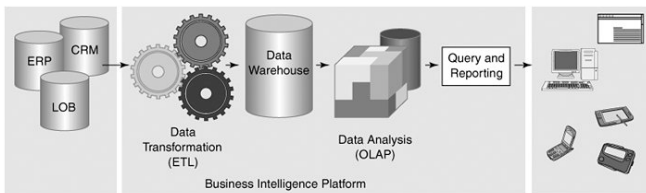

**Figure 1:** Business Intelligence Architecture (John C. Hancock, 2006)

## 1.2 BI example: the call detail record data mart

The example is for a service provider deciding to launch a marketing campaign dedicated for women on the occasion of the women's day. The purpose will be to offer women free calling hours. To be able to make the optimal strategic decisions, the service provider needs to answer the following question:

*What is the percentage of the call duration in a specific hour of the day by gender?*

Business Intelligence comes to provide an answer to such questions. Therefore, we opt for designing and building a call detail record (CDR) data mart and generating synthetic graphs.

**Design of the CDR data mart:**
The data mart should integrate data from the relevant data sources in order to understand the customer's calling behavior. Figure 2 shows the CDR data mart model.

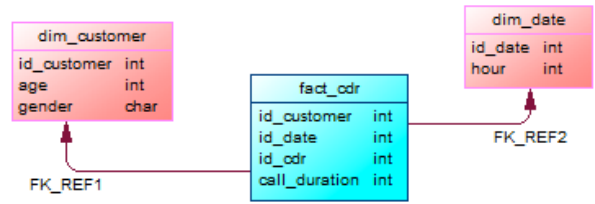

**Figure 2:** The CDR data mart star schema

We have executed the BI steps, previously explained, and we came out with the following reports:

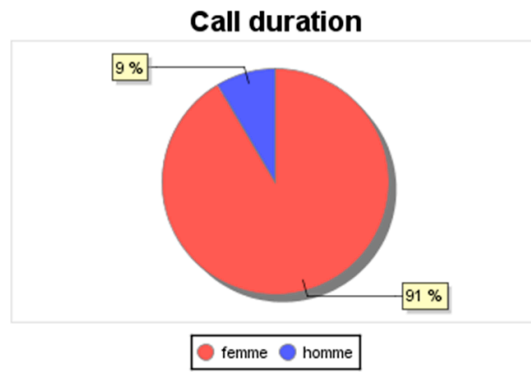

**Figure 3:** Graph - Call duration by gender at 4 pm

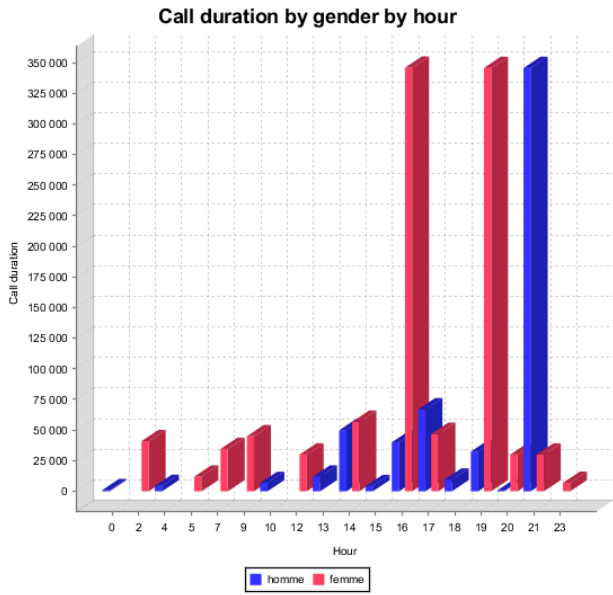

**Figure 4:** Bar Chart - Call duration by gender and by hour

Figure 3 and Figure 4 show that women are more likely to call at 4 pm and 7pm. Therefore, the best decision would be to plan the free calling hours outside of these hours, so that the service provider doesn't lose the profits usually gained at 4pm and 7pm.

## 2. Introduction to BI applied to Telecom

It is well known that the service providers dispose of a large amount of data though it is not always efficiently exploited. Providing frameworks for structuring the enterprise processes, information and measuring the Key Performance Indicators is one of the goals of the TM Forum community. The Information Framework along with the Business Metrics and the business process framework support the implementation of BI solutions within the telecommunication industry.

### 2.1 Use of eTOM, SID and Business Metrics

TM Forum has developed a set of Frameworks, called together Frameworx, giving a 360° view of the business since they address and cover its different key aspects. Each Framework has a specific focus:
- The Business Process Framework (also known as eTOM) has a Process Focus,
- The Information Framework (also known as SID) has an Information Focus,
- The Application Framework (also known as TAM) has an Application Focus,
- The Integration Framework is focused on the integration issues and challenges,
- The Business Metrics is focused on how to measure the performance of the business based on a balanced scorecard that covers:
    o The Customer Experience
    o The Operational Efficiency, and
    o The Revenue and Margin.

The above Frameworks are interrelated. For example, The Business Process Framework defines the processes, SID defines the information used and acted on by the processes, and the Business Metrics Framework measures the performance of the processes.

In this project we have worked with eTOM, SID and Business Metrics.

### 2.2 eTOM fundamentals

#### 2.2.1 eTOM Introduction

The Business Process Framework is one component of the Frameworx (formerly known as NGOSS) Frameworks used to support the previously mentioned TM Forum mission with a process focus.
The purpose of the framework is to continue to set a vision for the industry to compete successfully through the implementation of business process driven approaches to managing the enterprise. This includes ensuring integration among all vital enterprise support systems concerned with service delivery and support. The focus of the framework is on the business processes used by organizations, the linkages between these processes, the identification of interfaces, and the use of Customer, Service, Resource, Supplier/Partner and other information by multiple processes. Exploitation of information from every corner of the business will be essential to success in the future. In an ebusiness environment, automation to gain productivity enhancement, increased revenue and better customer relationships is vital. Perhaps at no other time has process automation been so critical to success in the marketplace. The over-arching objective of the framework is to continue to build on TM Forum's success in establishing:

- An 'industry standard' business process framework.
- Common definitions to describe process elements of an organization.
- Agreement on the basic information required to perform each process element within a business activity, and use of this for business requirements and information model development that can guide industry agreement on interfaces, shared data model elements, and supporting system infrastructure and products.
- A process framework for identifying which processes and interfaces are in most need of integration and automation, and most dependent on industry agreement.

The Business Process Framework and the associated business process modeling, describes for an enterprise the process elements and their relationship that are involved in information and communications services and technologies management. Additionally, the points of interconnection that make up the end-to-end, customer operations process flows for Fulfillment, Assurance, Billing & Revenue Management within Operations, and for Strategy, Infrastructure & Product are addressed by the framework.

Organizations need this common framework of processes to enable them to do business efficiently and effectively with other entities and to enable the development and use of third-party software without the need for major customization. In an ebusiness environment, this common understanding of process is critical to managing the more complex business relationships of today's information and communications services marketplace. ebusiness integration among enterprises seems to be most successful through strong process integration. Recent industry fallout, particularly in relation to dotcoms, does not reduce the pressure for ebusiness automation – it strengthens the need to capitalize on ebusiness opportunities to be successful.

However, the framework is not just an ecommerce or ebusiness process framework, it supports traditional business processes with the integration of ebusiness.

## 2.2.2 Structure of eTOM

The Business Process Framework is a reference framework or model for categorizing all the business activities that an organization will use. It is not an organization business model. In other words, it does not constrain the strategic issues or questions, such as who an organization's target, customers should be, what market segments should the organization serve, what are an organization's vision, mission, and so forth. A business process framework is one part of the strategic business model and plan for an organization.

The framework is better regarded as a business process framework, rather than a business process model, since its aim is to categorize the process elements and business activities so that these can then be combined in many different ways, to implement end-to-end business processes (for example, fulfillment, assurance, billing & revenue management) which deliver value for the customer and the organization.

Organizations doing business in today's distributed value chain also need an industry standard framework for procuring software and equipment, as well as to interface with other organizations in an increasingly complex network of business relationships. Many organizations have contributed their own process models because they recognize the need to have a broader industry framework that doesn't just address operations or traditional business processes.

The TM Forum initially identified business processes as a consensus tool for discussion and agreement among organizations involved in the information, communications, and entertainment industry. This encouraged convergence and general support for a broad common base in this area, which has been built on and extended with the framework, to enable:

- Focused work to be carried out in TM Forum teams to define detailed business requirements, information agreements, business application contracts and shared data model specifications (exchanges between applications or systems) and to review these outputs for consistency
- Relating business needs to available or required standards
- A common process view for equipment suppliers, applications builders and integrators to build management systems by combining third party and in-house developments

The anticipated result is that the products purchased by organizations and network operators for business and operational management of their networks, information technologies and services will integrate better into their environment, enabling the cost benefits of end-to-end automation. Furthermore, a common industry view on processes and information facilitates operator-to-operator, operator-to-customer, and operator-to-supplier/partner process interconnection, which is essential for rapid service provisioning and problem handling in a competitive global environment. This process interconnection is the key to ebusiness supply chain management in particular.

The framework also provides the definition of common terms concerning enterprise processes, sub-processes and the activities performed within each. Common terminology makes it easier for organizations to negotiate with customers, third party suppliers, and other organizations.

The framework focuses on seven key concepts, or entities, or domains, with which an enterprise's processes act on and use as it carries out its mission. The seven concepts are shown in Figure 5.

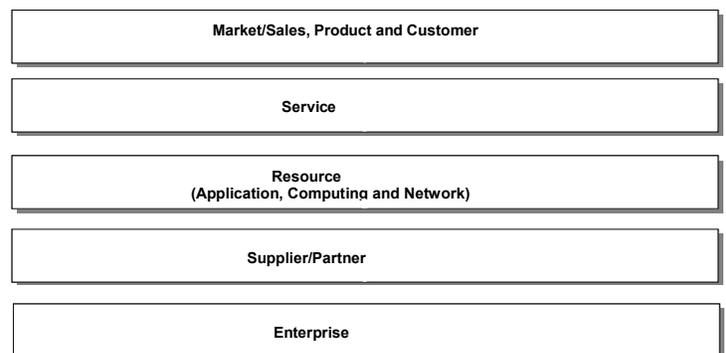

**Figure 5 Key Framework Concepts**

**Market/Sales** supports the sales and marketing activities needed to gain business from customers and potential customers. On the Sales side, this includes sales contacts/leads/prospects through to the sales-force and sales statistics. Market includes market strategy and plans, market segments, competitors and their products, through to campaign formulation.

**Product** is concerned with the lifecycle of products and information related to products' lifecycle. It includes the strategic portfolio plans, products offered, product performance, product usage, as well as the product instances delivered to a customer

**Customer** is individuals or organizations that obtain products from an enterprise, such as an organization. It represents of all types of contact with the customer, the management of the relationship, and the administration of customer data. Customer also includes data and related to the customer bills for products, collection of payment, overdue accounts, and the billing inquiries and adjustments made as a result of inquiries.

**Service** consists of information used to manage the definition, development, and operational aspects of services provided by enterprise. Service support various The Business Process Framework processes that deal with the definition, development and management of services used to realize products offered by an enterprise. This includes agreement on service levels to be offered, deployment and configuration of services, management of problems in service installation, deployment, usage, or performance, and quality analysis. Service also supports planning for future services, service enhancement or retirement, and capacity.

**Resource** consists of information used to manage the definition, development, and operational aspects of networks, as well as information and application resources that enable Products to be realized. It supports the The Business Process Framework processes that deal with the definition, development and management of the infrastructure of an enterprise. Resources also provide usage information which is subsequently aggregated to the customer level for billing & revenue management purposes. It also enables strategy and planning processes to be defined.

**Supplier/Partner** encompasses, planning of strategies for Supplier/Partners, handling of all types of contact with the Supplier/Partner, the management of the relationship, and the administration of Supplier/Partner data. It also supports bills, disputes and inquiries associated with a Supplier/Partner.

**Enterprise** represents information necessary to support the overall business, corporation or firm, which is using the Business Process Framework for modeling its business processes.

Developing a process framework starts with identifying key process areas that deal with these entities shown in Figure 6. These three areas represent the foundation of the business process framework and are referred to as Level 0 processes. Some guidance was provided by an earlier version of the process framework that focused only on operational processes, such as fulfillment and assurance. This process area is called Operations.

Considering that there must be a set of processes that enable operational processes, a second key process area can be defined. This process area deals with strategy, infrastructure, and lifecycle processes upon which operational processes depend. Included in this process are processes that define market strategies, resource strategies, and the offerings that are made available to the market. It is referred to as the Strategy, Infrastructure, and Product process area.

The third major process area, concerned with management of the enterprise itself, includes processes that support the other two areas. The idea for this process area comes from other enterprise-wide process frameworks, such as Michael Porter's value chain that includes a number of supporting processes, such as Human Resources. This process area is called Enterprise Management.

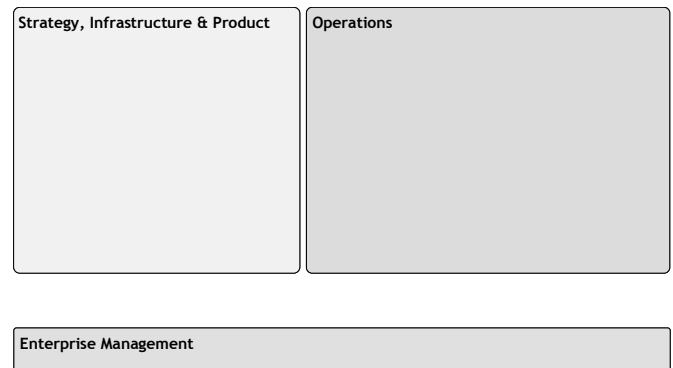

**Figure 6 Level 0 Process Areas**

One final aspect of the overall process framework shows the internal and external entities that interact with the enterprise. These are shown as (as ovals) in Figure 7. This figure, which is sometimes referred to as the Business Process Framework level 0 diagram, also shows the key business entities overlaid upon the level 0 processes that act upon them.

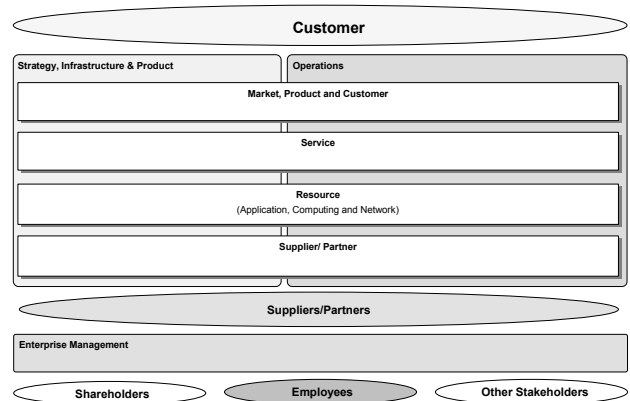

**Figure 7 The Level 0 Conceptual Process Framework**

### 2.2.3 Decomposing eTOM

The Business Process Framework represents a process hierarchy. Each level of the hierarchy represents the decomposition of the higher level. For details related to eTOM decomposition, you can refer to the article named

"Design of an enhanced Integrated Management System with Customer Experience Focus: The Business Process Framework (also known as eTOM) and ISO9001 together".

## 2.3 SID fundamentals

The Information Framework, also known as SID, provides a standard information structure, terminology, decomposition hierarchy, and classification scheme. The basic information container is called Business Entity. Of course, the information structure will contain may Business Entities (BE (s)). A business entity (BE) is a thing of interest to the business. It might be tangible such as customer, conceptual such as customer account, or active such as customer order. A BE, called also Class in UML (Unified Modeling Language), is composed of a name, attributes, and methods which is the BE's behavior. These BE (s) are interrelated. These interrelationships are also called, in UML, associations. SID focuses on BE's names, attributes and associations. Whereas the BE's methods are given by Services, APIs, and processes. These BE(s) are grouped using affinity analysis. The resulting groupings are called Aggregate Business Entities (ABE (s)). These ABE (s) could also be grouped. The highest level of ABE (s) groupings is called Level1 ABE(s). A domain is a collection of Level1 ABE(s) belonging to a management area. For these management areas, we can refer to the seven eTOM concepts which are Market/sales, Product, Customer, Service, Resource, Supplier/Partner, and Enterprise. SID has an additional specific domain called Common Business Entities (CBE) domain. This domain contains two types of ABE (s). The first type is the ABE (s) that can be placed into two or more domains and since SID in non redundant, we put them in the CBE domain. The second type is about the generalized ABE(s) such as Performance ABE.

Erreur ! Source du renvoi introuvable. shows the Level1 aggregate business entities and the domains to which they belong.

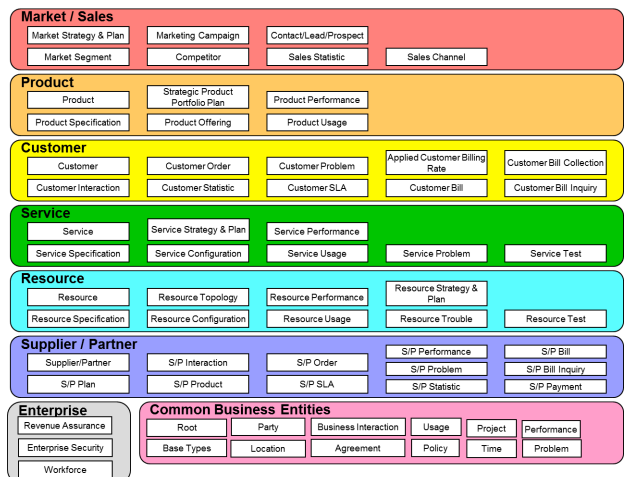

Figure 8 Information Framework Domains and their related Level 1 ABEs, Source: TM Forum

## 2.3.1 Extending the SID Models: adding attributes [1]

There are guidelines defined by TM Forum that should be used when extending SID. The guidelines include:
- Creating packages to hold Information framework extensions.
- Adding attributes.
- Adding new entities.
- Adding associations.

For the Purpose of the project we will only specify the guidelines for adding attributes.

As per TM Forum, there are four techniques that can be used to add attributes to an existing entity. As per TM Forum, There are four techniques that can be used to add attributes to an existing entity. If an entity is not sub-classed, then create a subclass of the Information Framework model business entity to which the attributes will be added. The subclass will inherit all of the attributes and associations from the Information Framework business entity thus maintaining the integrity of the Information Framework model. The new subclass holds the attributes as shown in Figure 9 below.

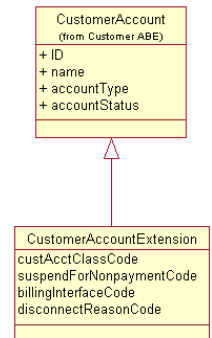

Figure 9 Adding Attributes Using Sub-Classing, Source: TM Forum

The name of the business entity that holds attribute extensions is an example. The actual name is at the discretion of the individual extending the Information Framework model. At a minimum, a consistent naming convention should be used.

Another technique can be used when the entity to be extended is already sub-classed. Here the attributes are "inherited" via the association, often referred to as an "IsA" association, implying the extension is a type of ServiceSpecification in Figure 10.

---

[1] Source : TM Forum

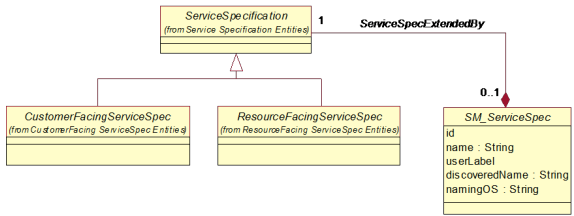

**Figure 10 Adding Attributes Using Composition, Source: TM Forum**

The composition type of association is used to indicate that the life cycle of the framework entity and the entity containing the enterprise specific attributes share the same lifecycle.  Where the composition appears in the association is up to the modeler.  Here the composition is from the extension to the framework entity implying the key entity is the extension.  If the modeler wishes to keep the framework entity as the key entity, then the composition would be from the framework entity.  Some modelers prefer to use an aggregation type of association, which is also acceptable. This is the most stable technique for adding attributes to an existing entity.

The third technique can be employed when there are a large number of extensions to make.  For example, one member added extensions to half the framework's entities, which would have increased the number of entities in the extended framework from about 1200 to 1800, increasing the apparent complexity of the model and making it more of a challenge to manage.

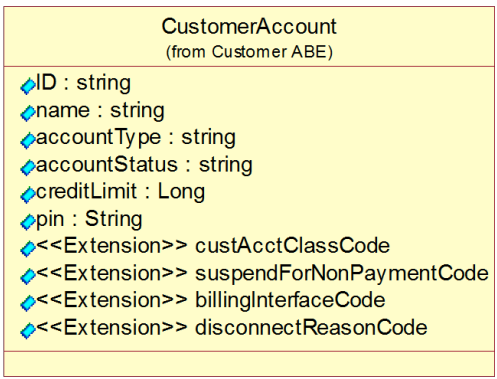

**Figure 11 Adding Attributes Using Stereotypes, Source: TM Forum**

The disadvantage of this technique is that version control would have to be manually performed.  For example, if a new version of CustomerAccount is introduced in the framework, either the extensions would have to be re-added or the new framework attributes would have to be manually added.  Most framework entities are stable from their initial introduction, so the disadvantage of this technique rarely comes into play.

The fourth technique that can be employed is using the generalized CharacteristicSpecification / CharacteristicValue pattern that enables the dynamic addition of attributes.

## 2.4 Business Metrics fundamentals

The Business Metrics, as per TM Forum, are based on a scaffold that targets business performance based on a balanced scorecard.

**Erreur ! Source du renvoi introuvable.**2 presents the three areas or domains of the balanced Scorecard:

- Revenue and Margin – financial performance indicators
- Customer Experience – indicators from the customer-facing side of the business
- Operational Efficiency – indicators around the key operational process areas of fulfillment, assurance, billing, and Call Center.

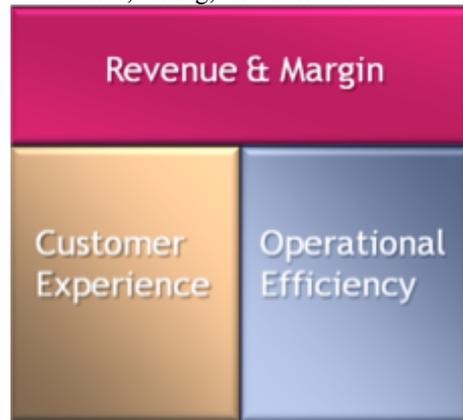

**Figure 12 the Balanced scorecard (TM Forum, BM1001 Metrics Development Guide)**

As shown in **Erreur ! Source du renvoi introuvable.**, the balanced scorecard suggests five process focus areas: General, Customer Management, Fulfillment, Assurance and Billing. The categorization enables to focus the measure on a specific process area.

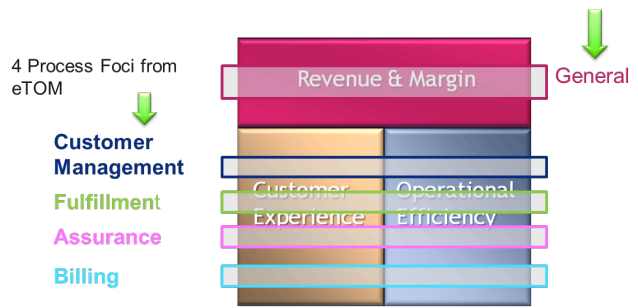

**Figure 13 Process Foci (TM Forum, BM1001 Metrics Development Guide)**

Each metric, as per TM Forum, has an area or domain, process focus and topic. The topics considered in the balanced scorecard areas are:

- Revenue And Margin

1. Margin/Revenue
   2. OpEx/CapEx
   3. OpEx/Revenue
   4. Revenue Breakdown
   5. Customer Constancy
- Customer Experience
   1. Preferred Access
   2. Customer Time Spent
   3. Usability
   4. Accuracy
   5. Contact Availability
   6. Ease of Doing Business
   7. Pricing Flexibility
   8. Security
- Operational Efficiency
   1. Unit Cost
   2. Time
   3. Rework
   4. Simplicity
   5. Process Flexibility
   6. Utilization

## 3. The Customer Experience Data Mart

The Customer Experience, as per TM Forum, can be defined as the result of the sum of observations, perceptions, thoughts and feelings arising from interactions and relationships (direct and indirect) over an interval of time between a customer and their provider(s) when using a service.

### 3.1 The Customer Experience Corner Stones

In addressing customer experience one needs to have a global view of the relationship between the customer and the provider. It is important to note that this relationship starts before delivering any service to the customer (**Pre-Service**) and evolves with the different experiences perceived during the operation of the service (**In-Service**)

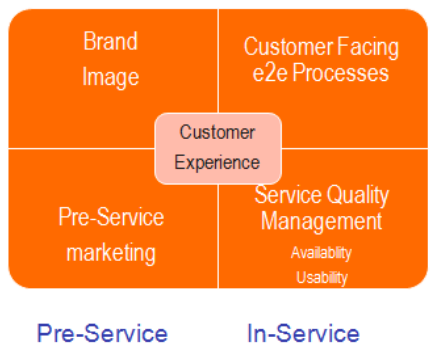

**Figure 14 Four Cornerstones of Customer Experience, Source: TM Forum**

**Erreur ! Source du renvoi introuvable.** shows the four corner stones of the customer experience. The details related to these corner stones, as per TM Forum, are:

- **Brand Image corner stone:** This pillar is concerned with ensuring that the corporation image leaves the best impression in the customer's minds, and developing this impression overtime through advertising campaigns of the company's values.
- **Pre-service Marketing corner stone:** This pillar is concerned with marketing campaigns that connect with the customers before they become customers, and presents the service offerings that meet their needs in order to encourage them to adopt the product.
- **In-service Customer Facing processes corner stone:** This pillar is concerned with the end to end customer facing services that support the customer centric end to end processes in Fulfillment, Assurance, and billing and revenue management. This particular corner Stone will be, in part, the focus of this study.
- **In-service Quality Management corner stone:** This pillar is concerned with ensuring the availability and usability of the service to the customers, and guarantying that the service quality meets their Pre-Service expectations and needs.

### 3.2 Customer Centric end to end processes detailed [1]

The customer centric Processes model the customer's interaction with the service provider. They start with the customer initiated contact and end with the fulfillment of the request, needless to say it is admitted that the satisfaction of the request's fulfillment will be measured.
A Total of seven customer centric processes are identified:

- **Request to Answer:** This process comprises activities relevant to managing customer requests across all communication interfaces. Furthermore, it addresses special requests that could lead to the preparation of a pre-sales offer.
- **Order to Payment:** This process deals with all activities that enable a customer's request or an accepted offer to become a ready-to-use product. This process is initiated by the customer contact, upon which the customer order information data is acquired then the appropriate process is triggered for the order handling. Once the service is delivered successfully, the customer order is closed and the customer's satisfaction is validated. Throughout this study we will focus particularly on this customer centric end to end process as a proof of concept.
- **Usage to Payment:** This process deals with all activities related to the handling of the product/service usage, including pricing, recording the data usage and generating the bill.
- **Request to change:** This process deals with the activities which convert the customer's change request into a ready-for-use product. In this regard, this process is a specification of order to payment, since it only deals with change requests.

---

[1] Source : TM Forum

- **Termination to confirmation:** This process deals with all activities related to the execution of the customer's termination request.
- **Problem to Solution:** This process deals with a technical complaint (problem) initiated by the customer, analyzes it to identify the root causes, initiates resolution, monitors progress and closes the trouble ticket. The basis of the problem is an unplanned interruption to a product/service or reduction in the quality of product/service.
- **Complaint to solution:** This process is a specialization of Problem to solution, it only deals with inquiries in which the customer expresses his/her dissatisfaction with a product or and inquiry handling speed.

## 3.3 Scope related to the article[1]: Order to Payment

As explained earlier, the Order to Payment process deals with all activities which convert the customer request or an accepted offer into a 'Ready for use' product. This process is composed of the following tasks:
- Handle customer contract
- Handle customer data
- Handle customer order
- Check creditworthiness
- Monitor the order
- Check order entry
- Initiate the production order
- Convert the customer interaction
- Consider service/resource/supplier partner layer
- Test services and resources
- Activate the products
- Trigger to start data collection for billing
- Generate & provide invoice
- Trigger to start ongoing operation
- Split the order
- Trigger to perform cross/up selling activities

In order to accomplish these tasks the following **inputs** are required: accepted offer, contract, inventory information, customer data, product elements, their relations and constraints, suppliers, distributors, etc.

As an **outputted result** of these activities: an invoice, a confirmation of a ready to use product and order confirmation are issued, and if necessary the software, hardware, and firmware are delivered to the customer.

The order to Payment Scenario is shown in **Erreur ! Source du renvoi introuvable.** bellow, in the form of a level 3 eTOM process flow. This scenario summarizes the tasks discussed in the chapter above.

---
[1] Source : TM Forum

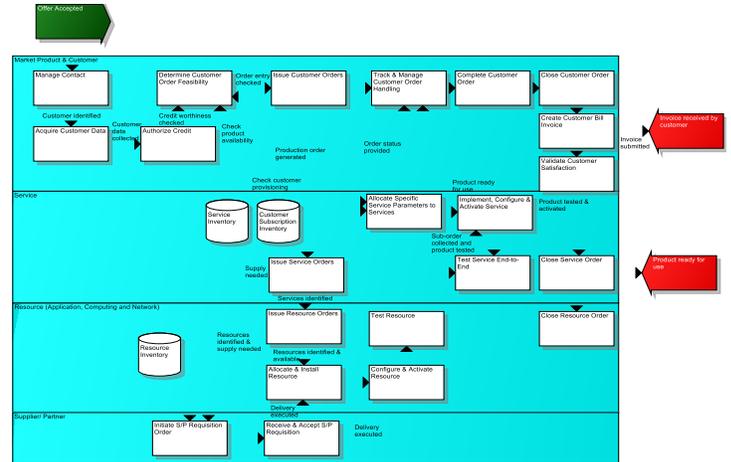

**Figure 15 Order to Payment - Process Flow (eTOM level 3 processes), Source: TM Forum**

In addressing the Order to payment process, we must take into account the following critical issues: the rework rate, the short cycle time between contract closure and service usage, In-time delivery product elements, time of order handling, Reliability, availability of the relevant services and resources, etc.

Defining these critical issues enables the selection of the appropriate business metrics for the performance measurement of this process.

## 4. BI and Telecom Methodology

The methodology used in the elaboration of this solution is both inspired from the iterative software development process: Rational Unified Process (RUP) and the GIMSI methodology for BI project management.

### 4.1 Rational Unified Process

RUP is an iterative, architecture centric approach. It's a well-defined and structured software engineering process, and it provides a customizable process framework.

The major milestones of the Rational Unified Process are:
- Inception Phase: this phase consists on understanding *what* to build
- Elaboration Phase: This phase consists on understanding *how* to build it
- Construction Phase: This phase consists on building the product
- Transition Phase: This phase consists on validating the solution

### 4.2 GIMSI methodology

GIMSI is a complete method to design and implement a decision support system. It is structured in 10 steps:
1. Identification of the strategy and the enterprise environment
2. Analysis of the enterprise architecture
3. Definition of the goals

4. Definition of the dashboard
5. Selection of the Key Performance Indicators matching the defined goals
6. Collecting the data necessary to the indicators construction
7. Construction of the dashboard
8. Choosing the adequate decisional software to implement the solution
9. Integration and deployment of the chosen software
10. Audit and verification of the functioning of the implemented decision support system

## 4.3 The Adopted Methodology

The adopted methodology is a combination between the RUP and GIMSI methodologies. As stated earlier, the idea behind the methodology adopted for this project, was to select the adequate aspects of each methodology that meet our needs. RUP strength is its iterative approach whereas the GIMSI strength is his focus on BI and decision making aspects. The suggested methodology is composed of the following phases:

### 4.3.1 Inception Phase

The main purpose of this phase is to perform process, information and business metrics high level and detailed scoping using the TM Forum documents and frameworks namely eTOM, SID and Business Metrics as they are. This phase is composed of:
Sub-Phase1: High Level Scoping
The related steps are:
1. High Level Process Scoping using eTOM Level2 process elements
2. High Level Information Scoping using SID ABE (s)
3. Key Performance Indicators (KPI (s)) Scoping using the Business Metrics Framework. These KPI (s) are the macro KPI (s) for the eTOM Level2 processes previously identified and for the eTOM Level3 processes which will be identified in the next sub-phase.
Sub-Phase2: Detailed Level Scoping
The related steps are:
1. Detailed Level Process Scoping using eTOM Level3 process elements
2. Detailed Level Information Scoping using SID BE(s)/attributes

### 4.3.2 Elaboration Phase

The main purpose of this phase is to analyze and refine the high level and detailed scoping previously done in the inception phase. In this phase, we will use cross-mapping between frameworks. These cross-mappings Process-Information-Business Metrics will help to confirm and elaborate the mappings done in the inception phase and they are perfectly aligned with the enterprise architecture mindset. The related steps to this phase are:

Sub-Phase1: High Level Scoping refinement and analysis
The related steps are:
1. High Level Process Scoping refinement
2. Mapping eTOM Level2 – Business Metrics and analysis
3. Mapping SID ABE (s) - Business Metrics and analysis
4. SID ABE (s) refinement

Sub-Phase1: Detailed Level Scoping refinement and analysis
The related steps are:
1. eTOM Level3 refinement
2. Mapping eTOM level 3 – Business Metrics and analysis
3. Mapping Business Metrics – SID BE(s)/attributes and analysis
4. SID BE(s)/ attributes refinement

### 4.3.3 Construction Phase

The aim of this phase is to build the data mart and implement it using one of the market tools which Pentaho in our case.
In order to model the Data mart, we need to specify the fact tables and the dimension tables.
As a starting point, we created a fact table for each selected metric (See Business Metrics Scoping). To determine the dimension tables, we used the output of the elaboration phase (See Mapping SID ABE (s) – Business Metrics).
After analysis, it was necessary to add the time dimension as an analysis axe for each fact table, because time is essential to any BI project.

### 4.3.4 Test Phase

The aim of this phase is to perform a BI end to end test from Data sources to reports. Since the whole a project is a research one, the data sources will be simulated using SID. SID will help to build the conceptual model. This model will be an input to generate the related logical and physical model.
An application of this methodology will be detailed in the coming chapters. The scope of this application is the customer end to end process "Order-To-Payment".

## 4.4 Inception Phase

### 4.4.1 High Level Scoping

#### 4.4.1.1 *High Level Process scoping using eTOM (level2)*

The key eTOM level 2, as per TM Forum documents, identified for the Order to Payment Process are:

- **Customer Interface Management**: this process manages all interfaces between the enterprise and potential and existing customers.
- **Selling**: is responsible for managing prospective customers, for the qualification and education of the customer and for matching customer expectations to the enterprise's products and services and ability to deliver
- **Order Handling**: is responsible for accepting and issuing orders. They deal with pre-order feasibility determination, credit authorization, order issuance, order status and tracking, customer update on order activities and customer notification on order completion.
- **Retention & Loyalty**: Manages all functionalities related to the retention of acquired customers, and the use of loyalty schemes in the potential acquisition of customers.
- **Bill invoice Management**: Ensures the bill invoice is created, physically and/or electronically produced and distributed to customers, and that the appropriate taxes, discounts, adjustments, rebates and credits for the products and services delivered to customers have been applied.
- **Service Configuration & Activation**: this process encompasses allocation, implementation, configuration, activation and testing of specific services to meet customer requirements, or in response to requests from other processes to alleviate specific service capacity shortfalls, availability concerns or failure conditions. Where included in the service provider offering, these processes extend to cover customer premises equipment.
- **Resource Provisioning:** this process is in charge of Allocating, installing, configuring, activating and testing specific resources to meet the service requirements, or in response to requests from other processes to alleviate specific resource capacity shortfalls, availability concerns or failure conditions.
- **S/P Requisition Management:** this process Tracks, monitors and reports on the service provider initiated product and/or service provisioning engagements to ensure that the interactions are in accordance with the agreed commercial arrangements between the service provider and the Supplier/Partner.

### 4.4.1.2 High Level Information scoping using SID ABE(s)

The key Aggregate business Entities, as per TM Forum documents, identified for the order to payment process are:
- **Customer:** Is the focus for the customer domain, Customer data is the enterprise's knowledge for the customer and accounts held by a customer.
- **Customer Order:** Handles single customer orders and the various types thereof, such as regulated and non-regulated orders.
- **Customer Interaction:** Represents communications with customers , and the translation of customer requests and inquiries into appropriate 'events' such as the creation of a customer bill inquiries, or the creation of a customer problem.
- **Customer Statistic:** Represents the analysis of customer usage patterns, customer profitability statistics and churn and retention statistics.
- **Customer Bill:** Handles real-time and non-real-time Call Detail Records (CDRs) and other sources of data that result in invoice items. The Customer Bill ABE also represents the format of a bill, schedule the production of bills, customer invoicing profiles, all the financial calculations necessary to determine the total of the bill, and credits and adjustments of bill.
- **Service:** Represents both customer-facing and resource-facing types of services. Entities in this ABE provide different views to examine, analyze, configure, monitor and repair Services of all types.
- **Service Order:** is a type of request that represents a customer Order's product decomposed into the services through which the product are realized.
- **Service Configuration:** Represents and manages configurations of **Customer Facing** Service and **Resource Facing Service** entities. It provides details on how the configuration can be changed. The entities depend on entities in the resource domain, which provide the physical and logical infrastructure for the implementing a Service.
- **Service Test:** Tests customer and resource facing entities. These entities are usually invoked during installation, as a part of trouble diagnosis or after trouble repair has been completed.
- **Resource:** Represent the various aspects of a Resource. This includes four sets of entities that represent: the physical and logical aspects of a Resource; show how to aggregate such resources into aggregate entities that have physical and logical characteristics and behavior; and show how to represent networks, sub-networks, network components, and other related aspects of a network.
- **Resource Order:** is a type of request that represents a Service Order's services decomposed into the resources on which the services will be provisioned.
- **Resource Configuration:** Represents and manages configurations of PhysicalResource, LogicalResource, and CompoundResource entities. It also provides details how the configuration is changed to meet product, service, and resource requirements, including activation, deactivation, and testing. Areas covered include verifying resource availability, reservation and allocation of resource instances, configuring and activating physical and logical resource instances, and testing.
- **Resource Test:** Tests Physical Resources, Logical Resources, Compound Resources, and Networks. These entities are usually invoked during installation, as a part of trouble diagnosis, or after trouble repair has been completed.
- **Supplier/Partner:** Is the focus for the Supplier-Partner Domain. Supplier-Partner represents the enterprise's knowledge of the Supplier Partner, their accounts and the relations the Enterprise has with the Supplier-Partner. It also contains all Supplier Partner agreements and negotiations.

### 4.4.1.3 KPI scoping using Business Metrics

The business Metrics, as per TM Forum documents, related to order to payment process are:

**F-CE-2a:** Mean duration to fulfill service order.
**F-CE-2b:** Mean time difference between customers requested delivery Date and Planned Date
**F-CE-2c:** % orders delivered by committed date
**F-CE-3:** % service usability queries
**F-CE-4:** % service activation failures
**F-CE-4b:** % of service faulty within 28 days of provisioning
**F-OE-2a:** Mean time order to activation
**F-OE-2b:** Order to activation time by major process
**F-OE-3a:** % orders requiring rework by cause type
**F-OE-3b:** Mean time to handle defects or rework from order to customer acceptance
**F-OE-3d:** %orders pending error fix

### 4.4.2 Detailed Scoping

### 4.4.2.1 Detailed level Process Scoping using eTOM (level3)

Table 1 lists the eTOM level2 processes as well as the related level 3 processes that appear in the Order-to-Payment process.

| eTOM Level2 Processes | eTOM Level3 processes |
|---|---|
| Customer Interface Management | Manage Contact |
| Selling | Acquire customer Data |
| Order Handling | Determine Customer Order feasibility |
| | Authorise Credit |
| | Track & Manage Customer order handling |
| | Complete customer order |
| | Issue customer orders |
| | Close customer Order |
| Retention & Loyalty | Validate customer Satisfaction |
| Bill invoice Management | Create Customer Bill invoice |
| Service Configuration & Activation | Allocate specific service parameters to services |
| | Implement & configure & activate service |
| | Test service end-to-end |
| | Issue service orders |
| | Close service order |
| Resource Provisioning | Allocate & install resource |
| | Configure & activate resource |
| | Test resource |
| | Issue resource orders |
| | Close resource order |
| S/P Requisition Management | Initiate S/P requisition order |
| | Receive & accept S/P requisition |

Table 1 eTOM level 3 processes, Source: TM Forum

### 4.4.2.2 Detailed Level Information scoping using SID BE(s)/attributes

| ABE name | Entity name | Attribute name |
|---|---|---|
| Customer ABE | Customer | ID |
| | | customerStatus |
| | | customerRank |
| | | partyRoleId |
| | | status |
| | | validFor |
| | | name |
| Customer order ABE | CustomerOrder | customerOrderType |
| | | purchaseOrderNumber |
| | | assignedPriority |
| | | assignedResponsibilityDate |
| | | dueDate |
| | | ID |
| | | interactionDate |
| | | description |
| | | interactionDateComplete |
| | | interactionStatus |
| | customerOrderItem | quantity |
| | | action |
| Customer Interaction ABE | CustomerInquiry | ID |
| | | interactionDate |
| | | description |
| | | interactionDateComplete |
| | | interactionStatus |
| Customer Statistic ABE | ChurnRetentionStatistic | |
| Customer Bill ABE | CustomerBill | billNo |

| | | billAmount |
|---|---|---|
| | | billDate |
| | | chargeDate |
| | | creditDate |
| | | mailingDate |
| | | paymentDueDate |
| | | relationshipType |
| Service ABE | Service | _partyRole2 |
| | | isServiceEnabled |
| | | hasStarted |
| | | isMandatory |
| | | startMode |
| | | isStateful |
| | | commonName |
| | | description |
| | | objectID |
| Service ABE::Service Order ABE | ServiceOrder | ID |
| | | interactionDate |
| | | description |
| | | interactionDateComplete |
| | | interactionStatus |
| | ServiceOrderItem | quantity |
| | | action |
| Service Test ABE | | |
| Service Configuration ABE | | |
| Resource ABE | Resource | usageState |
| | | managementMethodCurrent |
| | | managementMethodSupported |
| | | version |
| | | commonName |
| | | description |
| | | objectID |
| Resource ABE::Resource Order ABE | ResourceOrder | ID |
| | | interactionDate |
| | | description |
| | | interactionDateComplete |
| | | interactionStatus |
| | ResourceOrderItem | quantity |
| | | action |
| Resource Test ABE | | |
| Resource Configuration ABE | | |
| Supplier/Partner Order ABE | | |

**Table 2 SID BE / attributes, Source: TM Forum**

## 4.5 Elaboration phase

### 4.5.1 High level scoping refinement and analysis

#### 4.5.1.1 High Level Process scoping refinement

The results of the high level process scoping refinement are reported in Table 3.

#### 4.5.1.2 Mapping eTOM Level2 – Business Metrics and analysis

In Table 3, we present also the results of the mapping between eTOM Level2 – Business Metrics and analysis. Considering the example of F-CE-2a KPI, the process scoping was enhanced by adding 'Order Handling' process.

*NB: for all the tables presented in this phase, the cells highlighted in blue indicate the updates added to the standard in the first iteration, yellow refers to the second iteration refinements.*

| eTOM level 2 Process | CE | | | | | | OF | | | | |
|---|---|---|---|---|---|---|---|---|---|---|---|
| | F-CE-2a | F-CE-2b | F-CE-2c | F-CE-3 | F-CE-4 | F-CE-4b | F-OE-2a | F-OE-2b | F-OE-3a | F-OE-3b | F-OE-3d |
| Customer Interface Management | | | | √ | √ | √ | | | | | |
| Order Handling | √ | √ | √ | | √ | | √ | √ | | √ | √ |
| Retention & Loyalty | √ | | √ | √ | √ | √ | √ | √ | √ | √ | √ |
| Bill invoice Management | | | | | | | | | | | |
| Service Configuration & Activation | | | | √ | √ | √ | √ | √ | √ | √ | |
| Resource Provisioning | | | | √ | √ | | √ | | | | |
| S/P Requisition Management | | | | | | | | √ | | | |

**Table 3 Mapping eTOM (level2) - Business Metrics, Source: TM Forum**

#### 4.5.1.3 Mapping SID ABE (s) - Business Metrics and analysis

Following the same logic, in the case of the F-CE-2a KPI, it was important to add 'CustomerInteraction', 'Customer'

and 'CustomerOrder' ABEs to the high level information scoping, since they are used in this metric's calculation. Likewise, from this reevaluation, "ServiceProblem", "TroubleTicket" and "Place" were added.

| SID ABE (s) | CE | | | | | | OE | | | | |
|---|---|---|---|---|---|---|---|---|---|---|---|
| | F-CE-2a | F-CE-2b | F-CE-2c | F-CE-3 | F-CE-4 | F-CE-4b | F-OE-2a | F-OE-2b | F-OE-3a | F-OE-3b | F-OE-3d |
| Customer Interaction | √ | | | √ | √ | √ | | | | | |
| Customer | √ | √ | √ | | √ | √ | | √ | | √ | |
| Customer Order | √ | √ | √ | | √ | √ | √ | √ | | √ | √ |
| Customer Statistic | √ | | √ | √ | | √ | √ | √ | √ | √ | √ |
| Service | | | | √ | √ | | | √ | | | |
| Service Order | | | | | | | √ | √ | √ | √ | |
| Service Configuration | | | | | | | | √ | | | |
| Service Test | | | | | | | | √ | | | |
| Resource | | | | | | | | √ | | | |
| Resource Order | | | | | | | | √ | | | |
| Resource Configuration | | | | | | | | √ | | | |
| Resource Test | | | | | | | | √ | | | |
| Supplier/ Partner Order | | | | | | | | √ | | | |
| ServiceProblem | | | | | √ | | | | | | |
| Troubleticket | | | | | | | √ | | | √ | √ |
| Place | | | | | | | √ | | | | |

**Table 4 Mapping SID ABE - Business Metrics**

Based on this mapping, we have restricted the SID ABE to be considered to the following list:

- **Customer Interaction**
- **Customer**
- **Customer Order**
- **Customer Statistic**
- **Service**
- **Service Order**
- **Service Configuration**
- **Service Test**
- **Resource**
- **Resource Order**
- **Resource Configuration**
- **Resource Test**
- **Supplier/ Partner Order**
- **Service Problem**
- **Trouble ticket**
- **Place**

### 4.5.1.4 SID ABE (s) refinement

The results for ABE (s) refinement are reported in the Table 4.

## 4.5.2 Detailed Scoping: refinement and Analysis

### 4.5.2.1 eTOM Level3 refinement

After going through each metric, we have removed the unused processes. In other words, the processes that do not appear in any metric, such as 'Acquire Customer Data' was eliminated.

### 4.5.2.2 Mapping eTOM level 3 – Business Metrics

In this phase of our research we have reevaluated the mapping between the eTOM Level3 and the business metrics that were presented in the previous chapters.

If we consider the same metric F-CE-2a, we are interested in *the sum of time from order placement to order acceptance by the customer*. Now, when going back to the Order-to-Payment process flow, a Customer order starts with an interaction handled by the 'Manage Request' process. Afterwards, there is a study of the Order feasibility; this is handled by the 'Determine Customer Order Feasibility' process. Then the Customer Order is issued, handled and closed. As for the 'Track and Manage Customer Order handling' process, in some cases the Order might require rework, therefore, it is re-opened. Consequently, adding this process in the mapping was mandatory. As far as 'Validate Customer Satisfaction' is concerned, the F-CE-2a metric is more focused on measuring the performance of the Order Handling processes, rather than validating the customer satisfaction. For this reason it was removed from the mapping. The same analysis was applied for the other metrics. The final result is also shown in the Table 5.

| eTOM Level3 processes | Fulfillment | | | | | | Operational Efficiency | | | | |
|---|---|---|---|---|---|---|---|---|---|---|---|
| | F-CE-2a | F-CE-2b | F-CE-2c | F-CE-3 | F-CE-4 | F-CE-4b | F-OE-2a | F-OE-2b | F-OE-3a | F-OE-3b | F-OE-3d |
| Manage Request | √ | | | √ | √ | √ | √ | | | | √ |
| Determine CO feasibility | √ | √ | √ | | | | √ | | √ | √ | |
| Track & Manage CO handling | √ | | √ | | √ | | √ | | √ | √ | √ |
| Issue CO | √ | | √ | | | | √ | √ | √ | √ | √ |
| Close CO | √ | | √ | | | | √ | √ | √ | √ | |
| Validate | | | | √ | √ | | √ | | | √ | √ |

| | | | | | | | | | |
|---|---|---|---|---|---|---|---|---|---|
| customer Satisfaction | | | | | | | | | |
| Implement & configure & activate service | | | √ | | √ | √ | √ | | |
| Test service end-to-end | | | √ | √ | √ | | | √ | √ | √ |
| Issue SO | | | √ | | | | √ | √ | | |
| Close SO | | | √ | | | √ | √ | √ | | |
| Allocate & install resource | | | | | √ | | √ | | | |
| Test resource | | | | √ | √ | | | | | √ |
| Issue RO | | | | | | | √ | | | |
| CloseRO | | | | | | | √ | | | |
| Initiale S/P requisition order | | | | | | | √ | | | |
| Receive & accept S/P requisition | | | | | | | √ | | | |

**Table 5 Final Mapping eTOM Level 3 - Business Metrics**

### 4.5.2.3 Mapping Business Metrics – SID BE(s)/attributes and analysis

In this phase we have studied each metric, identified the attributes used to calculate the numerator, the denominator and the parameters.

Table 6 shows the results of the mapping for the F-CE-2a KPI. The Numerator will be calculated by measuring for each **CustomerOrderID** the difference **[interactionDateComplete - interactionDate];** while the denominator, will be calculated by counting the **customerOrderID**s for which the **interactionDateComplete** is included in the **reporting period**. The Parameters gives an opportunity to further segment the data to be provided. The attributes highlighted in yellow are extensions to the Information Framework, using the SID extension guidelines previously described. These extensions are essential to the calculation of the related KPI. The same analysis was applied to the remaining metrics.

| F-CE-2a | | |
|---|---|---|
| Mean duration to fulfill customer Order | | |
| Numerator | sum of time from order placement to customer acceptance for all orders | |
| | Customer Order | CustomerOrderID |
| | | interactionDate |
| | | interactionDateComplete |
| Denominator | number of completion accepted by the customer during the reporting period | |
| | Customerorder | CustomerOrderID |
| | | interactionDateComplete included in reprting period |
| Parameters | Customer | PartyRoleName |
| | time | reporting period |

**Table 6 Mapping F-CE-2a - SID attributes**

For F-CE-2a, The Numerator will be calculated by measuring for each CustomerOrderId, the difference :( InteractionDateComplete - InteractionDate)
The denominator will be calculated by counting the customerOrderIDs for which the interactionDateComplete is included in the reporting period.

| F-CE-2b | | |
|---|---|---|
| Mean time difference between customer requested delivery date and planned date | | |
| Numerator | sum of calendar time from CRD to CCD for all installation commitment in period (where CCD is later than CRD) | |
| | customer Order | CustomerOrderID |
| | | CustomerRequiredDate |
| | | dueDate in reporting period |
| Denominator | Total number of installation committment | |
| | customer Order | CustomerOrderID |
| | | dueDate in reporting period |
| Parameters | time | reporting period |
| | Customer | PartyRoleName |

**Table 7 Mapping F- CE- 2b - SID attributes**

ForF-CE-2b, The Numerator will be calculated by measuring for each CustomerOrderId, the difference :( dueDate - customerRequiredDate)
While,The denominator will be calculated by counting the customerOrderIDs for which the DueDate is included in the reporting period.

## Table 8 Mapping F-CE-2c - SID attributes

| F-CE-2c | | |
|---|---|---|
| percentage of orders delivered by committed date | | |
| Numerator | Number of service deliveries or activations fulffilled before and until committed date | |
| | customer Order | CustomerOrderID |
| | | DueDate |
| | | DeliveryDate<=DueDate |
| Denominator | Total number of activations or deliveries in the reporting period | |
| | customer Order | CustomerOrderID |
| | | deliveryDate included in reporting period |
| Parameters | time | reporting period |
| | Customer | PartyRoleName |

For F-CE-2c, The Numerator will be calculated by counting the number of CustomerOrderId, for which the difference (dueDate-deliveryDate)>=0

While, the denominator will be calculated by counting the customerOrderIDs for which the deliveryDate is included in the reporting period

## Table 9 Mapping F-CE-3 - SID attributes

| F-CE-3 | | |
|---|---|---|
| percentage of Service Usability Queries | | |
| Numerator | number of customer Contacts relating to usability of service that has been installed | |
| | Customer Inquiry | CustomerInquiryID |
| | | InquiryType= usability inquiry |
| denominator | total number of service activations or deliveries | |
| | CustomerFacingService | CFSID |
| | | cfsStatus = 0 |
| Parameters | time | reporting period |
| | CustomerFacingService | serviceComponent |
| | Customer | PartyRoleName |

For F-CE-3, the Numerator will be calculated by counting the number of CustomerOrderId, for which inquiryType='usabilityInquiry'

The denominator will be calculated by counting the CFSIDs for which the cfsStatus=0

## Table 10 Mapping F-CE-4 - SID attributes

| F-CE-4 | | |
|---|---|---|
| percentage of service activation failures | | |
| Numerator | Number of service activations or deliveries that fail as a result f technical issues | |
| | ServiceProblem | ServiceProblemID |
| | | reason =' delivery or activation failure' |
| | | first alert=customer report |
| | Customer facing service | CFSID |
| | | cfsstatus=6 ( failed) |
| denominator | total number of completed service activation or deliveries | |
| | CustomerFacingService | CFSID |
| | | cfsStatus = 0 |
| Parameters | Customer | PartyRoleName |
| | time | reporting period |
| | CustomerFacingService | serviceComponent |

For F-CE-4, The Numerator is the sum of (ServiceProblemIDs for which reason='delivery or activation Failure') and (ServiceProblemIDs for which first alert='Customer Report') and (CFSIDs for which cfsStatus=6)

The denominator will be calculated by counting the CFSIDs for which the cfsStatus=0

## Table 11 Mapping F-CE-4b - SID Attributes

| F-CE-4b | | |
|---|---|---|
| percentage of service faulty within the first 28 days of service | | |
| Numerator | Number of customer Accepted Orders that fail within the first 28 days of service | |
| | customer Order | CustomerOrderID |
| | | deliveryDate |
| | ServiceProblem | TimeRaised |
| denominator | total number of accepted orders accepted | |
| | customer Order | CustomerOrderID |
| | | interactionDateComplete (included in the reporting period) |
| parameters | Customer | PartyRoleName |
| | CustomerFacingService | serviceComponent |
| | time | reporting period |

For F-CE-4b, The Numerator will be calculated by counting the number of CustomerOrderId, for which the difference ( TimeRaised - deliveryDate) <= 28 days

The denominator will be calculated by counting the CustomerOrderIDs for which interactionDateComplete is included in the reporting Period

| F-OE-2a | | |
|---|---|---|
| | mean time order to activation | |
| Numerator | Agregate time from order Date to Customer acceptance closed for this SO | |
| | Customer Order | BusinessInteractionID |
| | | interactionDate |
| | | interactionDateComplete |
| | Service Order | BusinessInteractionID |
| | | interactionDate |
| | | interactionDateComplete |
| | Resource Order | BusinessInteractionID |
| | | interactionDate |
| | | interactionDateComplete |
| Denominator | Number of Completions accepted by the customer during the reporting period | |
| | Customer Order | CustomerOrderID |
| | | Deliverydate |
| | | interactionDateComplete ( included in reprting period) |
| Parameters | Place | GeographicArea |
| | time | reporting period |

Table 12 Mapping F- OE- 2a – SID Attributes

Regarding F-OE-2a, for MetaProcess 1: for each BusinessInteractionID we will calculate the difference (ServiceOrder.interactionDate-CustomerOrder.InteractionDate)

For MetaProcess 2: for each BusinessInteractionID we will calculate the difference (ResourceOrder.interactionDate-ServiceOrder.InteractionDate)

For MetaProcess 3: for each BusinessInteractionID we will calculate the difference (ResourceOrder.interactionDateComplete-ResourceOrder.InteractionDate)

For MetaProcess 4: for each BusinessInteractionID we will calculate the difference (ServiceOrder.interactionDateComplete-ResourceOrder.InteractionDateComplete)

For MetaProcess 5: for each BusinessInteractionID we will calculate the difference (CustomerOrder.interactionDateComplete-ServiceOrder.InteractionDateComplete)

The denominator will be calculated by counting the customerOrderIDs for which the deliveryDate is included in the reporting period

| F-OE-2b | |
|---|---|
| order to activation time by major process | |
| Numerator | overall time from order to activation |

| | by process block(fulfillment meta process blocks 2-5) | |
|---|---|---|
| | Customer Order | BusinessInteractionID |
| | | interactionDate |
| | | interactionDateComplete |
| | Service Order | BusinessInteractionID |
| | | interactionDate |
| | | interactionDateComplete |
| | Resource Order | BusinessInteractionID |
| | | interactionDate |
| | | interactionDateComplete |
| Denominator | Number of Completions accepted by the customer during the reporting period | |
| | Customer Order | CustomerOrderID |
| | | DeliveryDate |
| | | interactionDateComplete included in reprting period |
| Parameters | Customer | PartyRoleName |
| | time | reporting period |

Table 13 Mapping F-OE-2b - SID Attributes

Regarding F-OE-2b, for MetaProcess 1: for each BusinessInteractionID we will calculate the difference (ServiceOrder.interactionDate-CustomerOrder.InteractionDate)

For MetaProcess 2: for each BusinessInteractionID we will calculate the difference (ResourceOrder.interactionDate-ServiceOrder.InteractionDate)

For MetaProcess 3: for each BusinessInteractionID we will calculate the difference (ResourceOrder.interactionDateComplete-ResourceOrder.InteractionDate)

For MetaProcess 4: for each BusinessInteractionID we will calculate the difference (ServiceOrder.interactionDateComplete-ResourceOrder.InteractionDateComplete)

For MetaProcess 5: for each BusinessInteractionID we will calculate the difference (CustomerOrder.interactionDateComplete-ServiceOrder.InteractionDateComplete)

The denominator will be calculated by counting the customerOrderIDs for which the deliveryDate is included in the reporting period

| F-OE-3a | |
|---|---|
| percentage orders requiring rework by cause type | |
| Numerator | Number of SO orders requiring rework from order to activation |

|  | Service Order | serviceOrderID |
|---|---|---|
|  |  | reworkNo > 0 |
| Denominator | Total number of SO Orders | |
|  | Service Order | serviceOrderID |
|  |  | interactionDateComplete |
| Parameters | time | reporting period |
|  | ServiceProblem | OriginatingSystem |
|  | CustomerFacingService | service component |

**Table 14 Mapping F-OE-3a SID attributes**

For F-OE-3a, the numerator will be calculated by counting the (serviceOrderIDs for which the >0)

The denominator will be calculated by counting the ServiceOrderIDs for which the interactionDateComplete is included in the reporting period

| F-OE-3b | | |
|---|---|---|
| Mean time to handle defect or rework from order to customer acceptance | | |
| Numerator | Aggregate elapsed time to resolve defect and reworks during order to customer acceptance for this SO | |
|  | Service Order | BusinessInteractionID |
|  | trouble ticket | BusinessInteractionID |
|  |  | TroubleDetectionDate |
|  |  | ServiceRestoredDate |
| Denominator | Total number of defect and rework incident | |
|  | Service Order | serviceOrderID |
|  |  | reworkNo > 0 |
| Parameters | CustomerFacingService | service component |
|  | ServiceProblem | OriginatingSystem |
|  | time | reporting period |
|  | Customer | PartyRoleName |

**Table 15 Mapping F-OE-3b - SID attributes**

For F-OE-3b, the numerator will be calculated by measuring the time (ServiceRestoredDate-TroubleDetectionDate) for each TroubleTicketID where (ServiceOrder.BusinessInteractionID =TroubleTicket.BusinessInteractionID)

The denominator will be calculated by counting the ServiceOrderIDs for which the interactionDateComplete is included in the reporting period

| F-OE-3d | | |
|---|---|---|
| %order Pending error fix | | |
| Numerator | total number of fulfillment orders with pending error fix | |
|  | Customer Order | BusinessInteractionID |
|  | trouble ticket | BusinessInteractionID |
|  |  | TroubleTicketState=Pending |
| Denominator | Total Number of fulfillment orders with errors | |
|  | Customer Order | BusinessInteractionID |
|  | trouble ticket | BusinessInteractionID |
| Parameters | Customer | PartyRoleName |
|  | time | reporting period |
|  | CustomerFacingService | service component |

**Table 16 Mapping F-OE-3d SID attributes**

For F-OE-3d, the numerator will be calculated by counting the TroubleTicketIDs for which (CustomerOrder.BusinessInteractionID=TroubleTicket.BusinessInteractionID) and TroubleTicketState=Pending

The denominator will be calculated by counting the TroubleTicketIDs for which (CustomerOrder.BusinessInteractionID=TroubleTicket.BusinessInteractionID)

### 4.5.2.4 SID BE(s)/ attributes refinement

The previous chapter had as an objective to determine the final selection of the SID ABE along with the attributes. The results are shown in the table, the attributes in the highlighted cells were added to the Information Framework. Note that this final selection will be the input for the Data Mart design, since the selected SID Business Entities correspond to the analysis axes.

The attributes highlighted in the Table 17 were extended to the SID model using the guidelines described in the chapter **Erreur ! Source du renvoi introuvable.**.

| SID BE | attributes |
|---|---|
| Customer Order | customerOrderID<br>InteractionDate<br>InteractionDateComplete<br>DeliveryDate<br>DueDate<br>CustomerRequestedDate<br>ReworkNo |
| ServiceOrder | serviceOrderID<br>InteractionDate<br>InteractionDateComplete<br>DeliveryDate<br>DueDate<br>CustomerRequestedDate<br>ReworkNo |
| ResourceOrder | resourceOrderID<br>InteractionDate<br>InteractionDateComplete<br>DeliveryDate<br>DueDate<br>CustomerRequestedDate<br>ReworkNo |
| serviceProblem | serviceProblemID<br>originatingSystem<br>reason |

| customerFacingServices | cfsID |
| | sfsStatus |
| | isServiceEnabled |
| | serviced |
| | hasStarted |
| | ==serviceComponent== |
| troubleTicket | troubleTicketID |
| | troubleTicketState |
| | troubleDetectionDate |
| | serviceRestoredDate |
| | InteractionDate |
| | InteractionDateComplete |
| Place | placeID |
| | geographicalArea |
| customer | customerID |
| | partyRoleName |
| **time** | **timeID** |
| | **Day** |
| | **Month** |
| | **year** |

**Table 17 The final selection of SID BE /attributes**

## 4.6 Construction Phase

The aim of this phase is to build the class diagram, design the data mart and to execute the BI steps presented earlier, in the scope of the customer centric end-to-end Process Order-to-Payment, based on the results from the elaboration phase.

### 4.6.1 Build of the class diagram

The object of this phase is to build a database model for generating the necessary data sources to our project.

Using the results of the detailed scoping in the elaboration phase, we ended up with a set of business entities. These business entities were the base of our database model; nevertheless, it was necessary to explore the Information Framework in order to extract the existing associations and aggregations between these classes while considering the relationships to the Common Business Entities.

For this, we were inspired by the top down method explained in the "Package Implementing the SID: Practitioner's guide". This technique is used to consolidate SID classes hierarchy by explicitly moving attributes and relationships from abstract super-classes to concrete subclasses. For instance, if we consider the classes 'BusinessInteraction', 'Request' and 'CustomerOrder', highlighted in Red in Figure 18. 'CustomerOrder' inherits from 'Request', and 'Request' inherits from 'BusinessInteraction'. Accordingly, the attributes of 'BusinessInteraction' will merge to the attributes of 'Request', and finally the result will merge to the attributes of 'CustomerOrder'.

Figure 18, depictures the final class diagram adopted for this project. The relationships between classes were conserved from the SID class model.

From the class diagram, we have generated the corresponding conceptual, logical and physical model.

The data used for implementing the BI solution, was manually generated while taking into consideration the database consistency constraints. For instance, for a certain row, the field interactionDate cannot have a higher value than interactionDateComplete.

### 4.6.2 The Order to Payment Data Mart

In order to model the Order to Payment Data mart, we need to specify the fact tables and the dimension tables.

As a starting point, we created a fact table for each selected metric. To determine the dimension tables, we used the output of the elaboration phase.
After reevaluating, it was necessary to add the time dimension as an analysis axe for each fact table, because time is essential to any BI project.

**Star-Schema Example:**
As a dimensional model, rather than the snow-flacked schema, the star schema was the more adequate since there are no hierarchical dimensions

Figure 16, shows an example of a star schema. The fact table contains the identifiers of all the dimensions to which it is linked along with the key performance indicator.

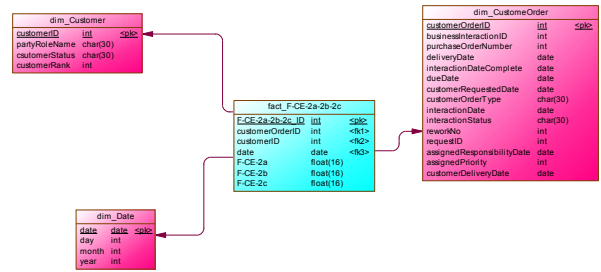

**Figure 16 F-CE-2a-2b-2c Star Schema**

**Constellation:**
After modeling all metric's data mart individually, the next step was to regroup all the metrics that share the same dimensions in one fact table in one constellation schema. This schema also allows the dimension tables to be shared amongst the fact tables, Figure 19, shows the Order to payment constellation diagram. This model contains 9 fact tables, the blue color indicates the fact tables related to the Customer Experiences KPIs, while the green indicates the fact tables related to the operational efficiency:
- In **Fact_F-CE-2a-2b-2c** we have regrouped the three KPIs F-CE-2a, F-CE-2b and F-CE-2c which

share the same dimensions. This fact table contains measures for these KPIs as well as the Dimension table's identifiers as foreign keys, namely the Customer, the Customer Order and the Time dimensions.

The remaining 8 fact tables have been modeled using the same principle described in the previous paragraph.

### 4.6.3 ETL Phase

After modeling the Data mart and having our data sources ready, the next step is to extract the information from the data sources, transform it and finally load it into the Data Mart.

For this step we have used the ETL tool Pentaho Data Integration, Figure 17 shows the loading of the dimension, this step didn't required any transformations since the data sources generated were at the right format.

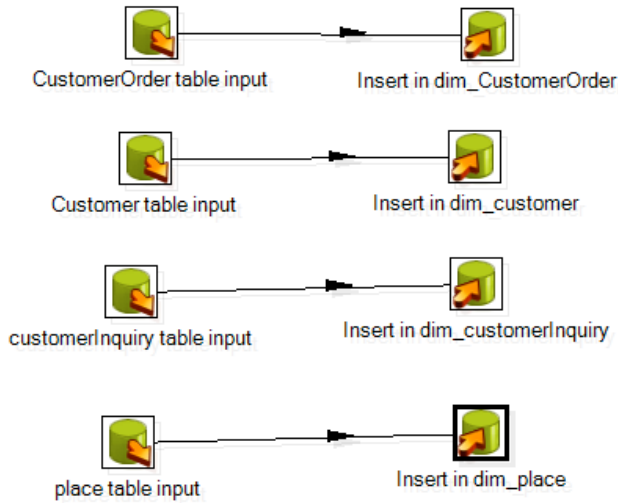

**Figure 17 Loading of the different dimensions**

Figure 20 shows the various transformations realized in order to calculate 'OrderDelay', 'OrderDuration' and 'OnTimeOrders' which are necessary to measure the KPIs of the fact table F-CE-2a-2b-2c. These transformations consist on joint table, selecting fields, filtering rows and calculating the measures.

For the three metrics F-CE-2a, F-CE-2b and F-CE-2c, the Customer Orders will be analyzed by 'Customer Segment' this information is contained in the Customer table. To be able to recreate the link between 'CustomerOrder', a businessInteraction, and Customer, a party Role, it was necessary to go through the table 'BusinessInteractionRole', which designates a Party or Resource playing a role in a BusinessInteraction, and the association Table 'BusinessInteractionRoleInvolvesPartyRole', as its name suggests the links between BusinessInterationRole and the PartyRole.

**Figure 18 the Order to Payment class diagram**

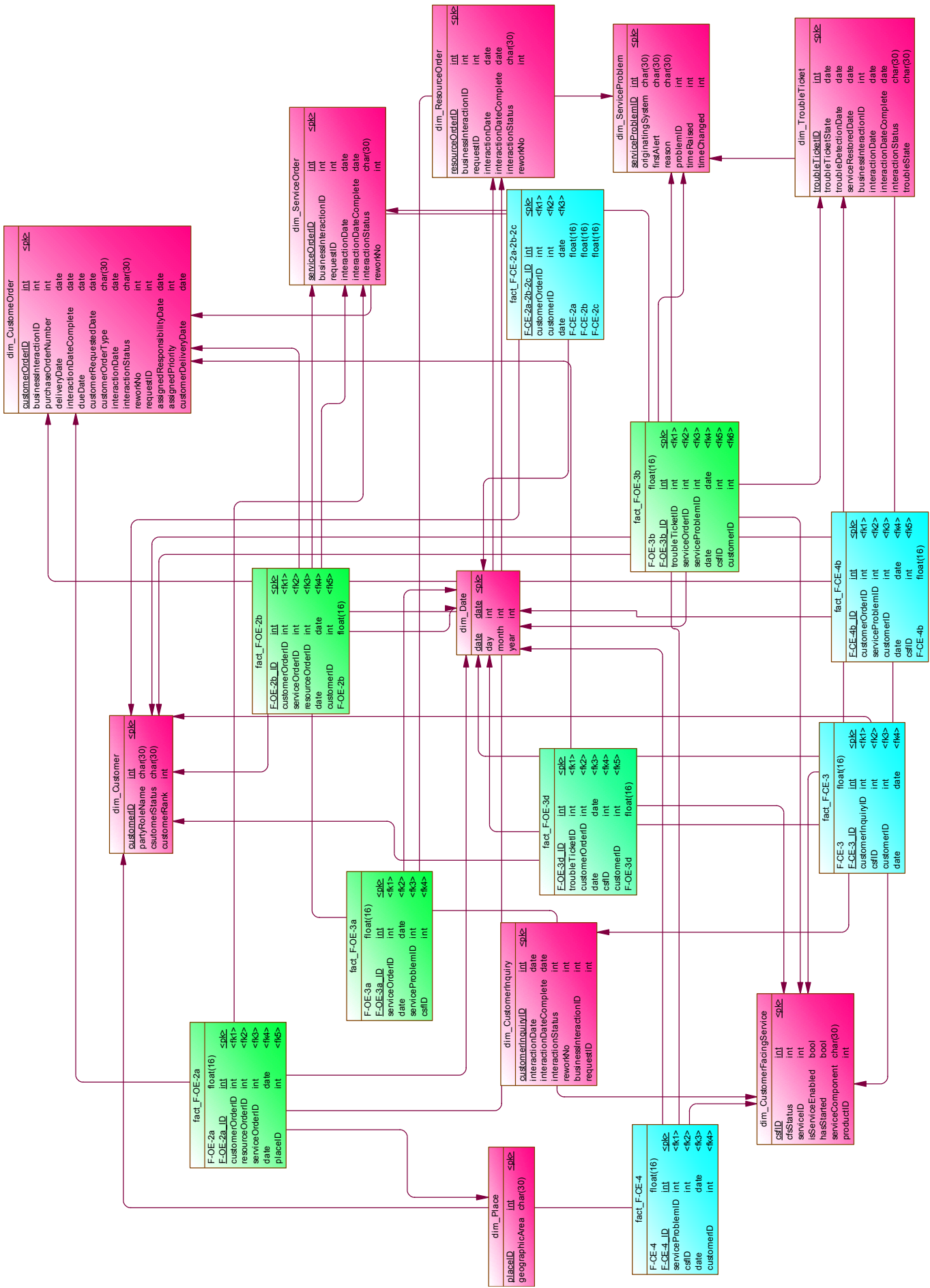

**Figure 19 The Order to Payment Data Mart**

This is why it was necessary to use 'Join on CustomerOrderId field', 'Join on BusinessInteractionRoleId field', and 'Join on CustomerOrderId'. At the end of each join step, it was essential to add a selection step, in order to filter the fields.

As for 'OntimeOrder calculation', it implied the difference between 'CustomerDeliveryDate' and 'DueDate'. The result of this step, is then filtered to conserve only the positive values that will be transformed into Boolean values since F-CE-2c requires the number of on time orders.

'OrderDuration' is the difference between 'InteractionDateComplete' and 'InteractionDate'. This calculation corresponds to F-CE-2a. It follows that for F-CE-2b, 'OrderDelay' is the difference between 'DueDate' and 'CustomerRequiredDate'.

The three calculations were then joined and loaded into the fact table F-CE-2a-2b-2c.

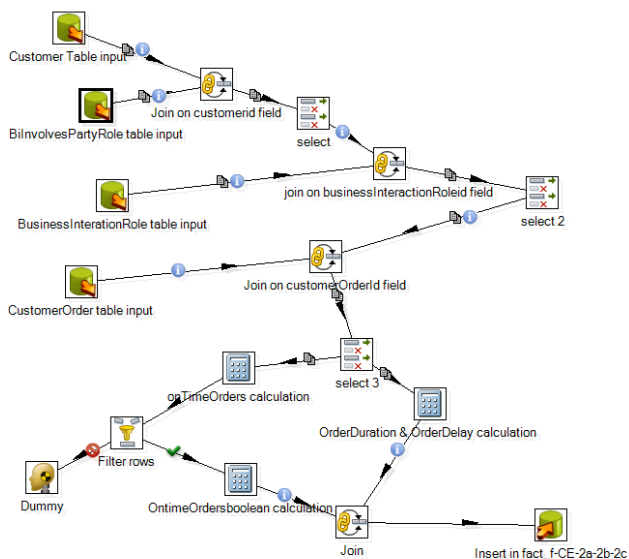

**Figure 20 Transformation of the fact table F-CE-2a-2b-2c**

### 4.6.4 Elaborating OLAP Cubes

After loading the Data Mart, the next step is to elaborate the OLAP cubes in order to analyze the selected KPIs considering the different analysis axis. For this purpose we used the Pentaho Analysis Services (PAS).

PAS consists of the following four components:
- **JPivot analysis front-end:** JPivot is Java-based analysis tool that serves as the actual user interface for working with OLAP cubes.
- **Mondrian ROLAP engine:** The engine receives MDX queries from front-end tools such as JPivot and responds by sending a multidimensional result-set.
- **Schema Workbench**: This is the visual tool for designing and testing Mondrian cube schemas. Mondrian uses this cube schemas to interpret MDX and translate it into SQL queries to retrieve the data from an RDBMS
- **Aggregate Designer:** is a visual tool for generating aggregate tables to speed up the performance of the analytical engine.

MDX is the abbreviation for Multi-Dimensional eXpressions, which is a language that is especially designed for querying OLAP databases. (Bouman & Dongen, 2009)

Figure 21 shows the OLAP cube created in Pentaho Schema Workbench for the multidimensional analysis of the F-CE-2a KPI.

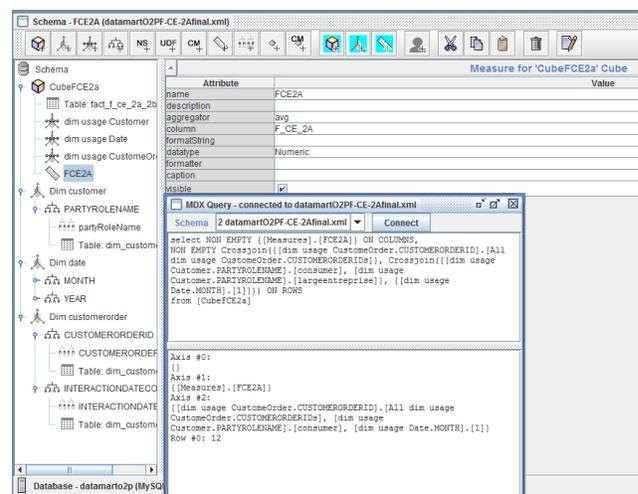

**Figure 21 the OLAP Cube for analyzing the F-CE-2a KPI**

### 4.6.5 Elaborate Report

The reports were created using the Pentaho Report designer tool and were published on Pentaho BI Server.

The report in Figure 22 shows the percentage of the mean time to fulfill Customer Orders by customer segments, for the month of January. This type of reports can give an idea to the decision makers about the customer orders processes efficiency. We can then detect and solve problems avoiding any bad customer experience proactively.

| Customer Segment | Order Duration (j) | Percent of total |
|---|---|---|
| consumer | 3,5 | 41,176% |
| largeentreprise | 5 | 58,824% |

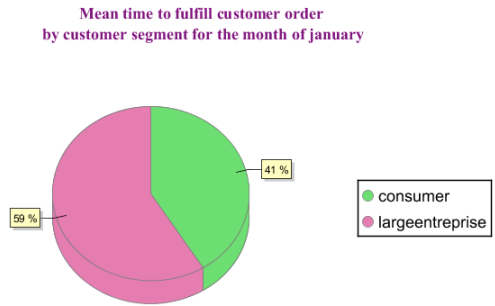

Figure 22: Chart Pie Report - Mean time to fulfill Customer Order by Customer segment for the month of January

The chart in Figure 23 illustrates the mean time to fulfill customer order for the large enterprise customer segment. In this report we analyze the customer order handling performance in depth, allowing us to monitor the evolution of this performance throughout each month of the year. The same analysis is presented in Figure 24, for the consumer customer segment.

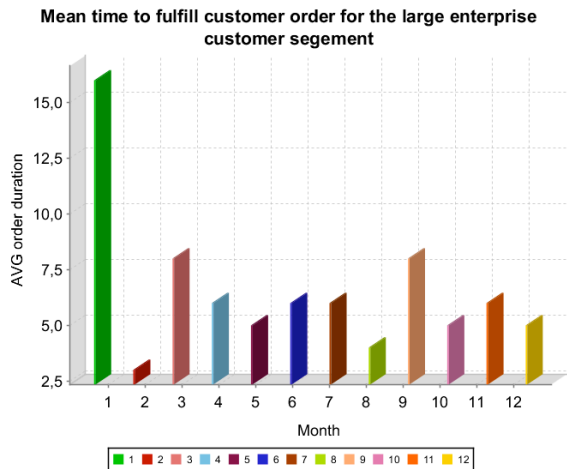

Figure 23: Bar Chart - Mean time to fulfill Customer Order for the large enterprise Customer Segment

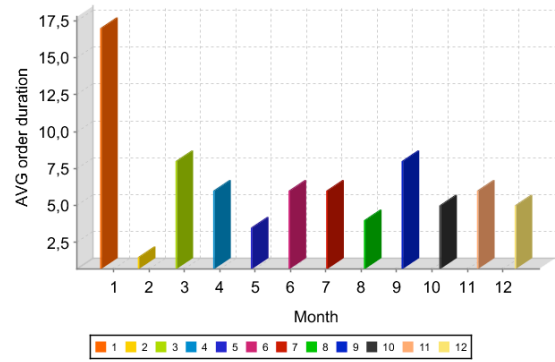

Figure 24: mean time to fulfill Customer Order for the consumer Customer Segment

## 5. Conclusion

This paper suggests a step by step methodology to design a Telco Data Mart and other BI steps. Order-To-Payment, an end to end customer process, was used as an application for this methodology. For this application the all the BI steps were considered. The findings of the current article could be also used to deploy a BI Test Bed in a Telco company. The main focus of this paper is the analysis done for the alignment of the three frameworks eTOM, SID and the Business Metrics. These different mappings enabled us to extend the SID framework with attributes which are required for the KPIs calculations.

Regarding the project's perspective, we suggest the reiteration of our methodology for the fifth remaining Customer Centric end-to-end processes, followed by the other Customer Experience Cornerstones, to cover all the aspects of the Quality of Experience. As for the Quality of Service, we suggest, the development of the Network Centric processes, using the same methodology.

**Acknowledgments**

Special thanks are addressed to our parents, our families and our friends for their prayers, guidance, smiles, advices and support.

Special thanks are addressed to TM Forum for its valuable input and support much appreciated.

The authors would like to thank Mr. Mike Kelly, TM Forum Senior Technical Manager, and Andrew Chalmers, TM Forum Director of Implementation Products, for their feedbacks, encouragements, and valuable inputs much appreciated.

We would like to thank in advance anybody for any provided feedback or for any inspiration.

**M. Benhima Mounire Benhima** is a Senior Consultant, specializing in Business Transformation with its implementation challenges. 15 years of experience in various industrial environments (mainly ICT) in many countries (Africa, America, Europe, Asia). He is distinguished to be the first one in the world to be certified The Business Process Framework Level4 (The highest The Business Process Framework certification level) with the highest score when it was an essay. He is also part of the international TM Forum Trainers Panel (21 members as per August 2012). Mounire has held various key consulting positions and has led numerous groups in major business transformation projects. As a result of his experience and background, Mounire has developed extensive operational and strategic skills in the Enterprise Architecture and Business Transformation with their impact on the organizational structure, quality management, business process engineering, information system governance, and change management.

**J. P. Reilly John P. Reilly** is member of the TM Forum's technical staff and a master trainer for the TM Forum. His technical staff responsibilities include providing Information Framework (SID) model and the Integration Framework subject matter expertise and program management in support of TM Forum programs and projects. In November 2005 John was elevated to Distinguished Fellow of the TM Forum in recognition of his contribution to the development of TM Forum Solution Frameworks (NGOSS) and the SID. He was instrumental in developing process and information models in support of the Forum's Revenue Assurance program. He is the author and co-author of a number of books and articles on these subjects.

**Z. Naamane** Zaineb Naamane is a Final year computer science engineering student at L'Ecole Mohammadia D'Ingénieurs, Rabat Morocco. Zaineb specializes in Information Systems and has a remarkable experience working with Business Intelligence software tools in designing and developing BI solutions. Zaineb has developed good team working skills during school projects and internships.

**M. Kharbat** Meriam Kharbat is a Final year computer science engineering student at L'Ecole Mohammadia D'Ingénieurs, Rabat Morocco. Meriam specializes in Software Quality and has an excellent Business Intelligence background acquired from the various projects held in the school and during internships.

**M.I Kabbaj** Mohammad. Issam Kabbaj Ph.D degree in Computer Science with distinction in 2009; Assistant Professor at the Computer Science Department-EMI; His interests include Software Process, Business Process, Risk Management, Process Mining, Dynamic Adaptation, MDA

**O. Esqalli** Oussama Esqalli is an IT / Business Process Management consultant, specializing in business transformation with a focus on processes and IT aspects, experience in various IT projects, he has a profile that combines the technical and functional aspects. He has contributed to the success of many missions of organization and business transformation; also he has led many projects of development and integration of software.